# Broadband and Accurate Electric Tuning of On-Chip Efficient Nonlinear Parametric Conversion

*Jiaqi Li[1,\*], Yanfeng Zhang[1,\*], Jinjie Zeng[1], and Siyuan Yu[1,\*]*

[1]State Key Laboratory of Optoelectronic Materials and Technologies, School of Electronics and Information Technology, Sun Yat-Sen University, Guangzhou 510006, China.

*Corresponding author: <u>zhangyf33@mail.sysu.edu.cn</u>; <u>yusy@mail.sysu.edu.cn</u>*

**Abstract**

On-chip nonlinear photonic conversion functions with wide and precise tunability as well as high conversion efficiency are highly desirable for a wide range of applications. Photonic crystal micro-ring resonators (PhCR) facilitate efficient nonlinear conversion and enable wavenumber-accurate selection of converted optical modes, but do not support post-fabrication reconfiguration of these operational modes. Coupled-ring resonators, on the other hand, allows post-fabrication reconfiguration but suffers from ambiguity in mode selectivity. We propose a novel segmented photonic crystal micro-ring resonator featuring half-circumference gratings that decouples the locking between the grating Bragg reflection peak and micro-ring resonance frequencies. By introducing complementary thermos-optical controllers that allow differential tuning between the grating reflection peak and the micro-ring resonance, the device supports electrically reconfigurable wavenumber-accurate optical mode selectivity, experimentally demonstrated as a voltage-tunable, power-efficient optical parametric oscillator. The device demonstrates electric tuning of signal and idler frequencies both in a per-FSR stepwise manner and in a gap-free continuous manner, achieving a broad optical frequency tuning range of > 5 THz and a conversion efficiency of >25%. The novel approach introduces unprecedented design flexibility as well as high and precise reconfigurability to integrated nonlinear photonics, providing a new pathway towards future high-performance on-chip nonlinear light sources.

**Introduction**

Micro-ring resonators (MRRs) are important components in integrated photonics, extensively used for linear signal processing[1] and nonlinear frequency conversion[2]. They strongly confine the energy of resonant optical modes to enhance their interaction with the optical medium, which allows for highly efficient operations in a compact desgin. Precise control of these modes is crucial for achieving desired functionalities. For instance, in MRR-based nonlinear conversion processes, the relative frequency detuning between resonant modes, i.e., the dispersion relationship of the MRRs, must be precisely controlled to meet specific phase-matching conditions[3,4].

Recently, wavenumber-accurate mode control using photonic crystal micro-ring resonators (PhCRs) has gained increasing attention in integrated nonlinear photonics[5–

[10]. By incorporating endless periodic gratings around the entire circumference of the micro-ring, the degeneracy of counter-propagating modes at the resonant frequency coinciding with the grating Bragg condition is lifted, resulting in distinct mode splitting and the introduction of a photonic bandgap in the dispersion profile[11]. This bandgap modifies the phase-matching conditions required for nonlinear conversion, enabling innovative applications such as spontaneous pulse formation[5,10], platicon microcomb initiation[6,9], and tailored wavelength outputs from Kerr micro-optical parametric oscillators (μOPOs)[7,8]. However, once fabricated, the grating structures embedded in the micro-ring cannot be tuned independently from the micro-ring resonances, leading to lower yields[5] and limited device functionality.

Alternatively, coupled micro-rings with two different free spectral range (FSR) offer broadband dispersion control through the Vernier effect, which periodically couples the two set of resonances. The physical separation between the auxiliary and main rings allows for realignment of their resonances through differential tuning, facilitating tunable μOPOs[12] and function-switchable light sources[13]. However, small variations in the FSR of the micro-rings result in significant uncertainty in the targeted mode numbers as well as their frequencies, especially pronounced with a much smaller auxiliary ring[14]. Moveover, unwanted nonlinear processes are not effectively suppressed using periodic broadband dispersion control, which limits the achievement of both high conversion efficiency and tunability in frequency convertors like μOPOs[12].

In this article, we propose a novel PhCR in which the operational modes can be accurately selected and dynamically controlled for precise dispersion management. The scheme comprise a micro-ring resonator embedded with a half-circumference grating segment that enables differential tuning of the Bragg reflection peak and the cavity resonances. Based on this new mechanism, we successfully fabricate and test an electrically tunable μOPO with a signal/idler tuning range exceeding 5 THz and a conversion efficiency surpassing 25%. Furthermore, we propose and demonstrate a complementary thermo-optic scheme that enables broadband, gap-free tuning of signal and idler frequencies with essentially fixed pump frequency, eliminating the need for an expensive, widely tunable pump laser source. Our new method can introduce unprecedented design flexibility and a high degree of reconfigurability to integrated nonlinear photonics platforms, including $\chi^{(3)}$ and $\chi^{(2)}$ media.

**Results**

**Scheme of electrically reconfigurable PhCR**

Figure 1 illustrates the scheme of an electrically reconfigurable PhCR, which comprises a MRR with a partial-circumference grating segment, termed a segmented PhCR, and a microheater positioned atop the grating region. The grating segment retains the accurate mode control capabilities of a full-circumference PhCR through the careful design of the grating period. When a mirco-ring resonant frequency $v_m$ coincides with the Bragg relfection frequency $v_B$, strong coupling between clockwise

(CW) and counterclockwise (CCW) propagating modes leads to the splitting of the original mode $v_m$ into two standing-wave modes $v_m^\pm = v_m \pm \beta_m$, with a frequency shift equal to the CW-CCW coupling rate $\beta_m$[11]. The coupling rate $\beta_m$ is affected by both the amplitude of the grating modulation and the alignment of the Bragg reflection peak with the target resonance $m$. Since the grating modulation amplitude is fixed post-fabrication, the reconfigurability of the PhCR relies on the independent tuning of the Bragg reflection peak $v_B$ relative to the micro-ring resonances $v_m$. In a full-circumference PhCR, however, the effective refractive index of the grating waveguide is uniform across the entire micro-ring and determines both the grating Bragg peak and the MRR resonance frequencies. The grating therefore always target the MRR resonance mode number $m_0$ coinciding with half the number of grating periods $N_0/2$. Therefore the PhCR has very limited reconfigurability.

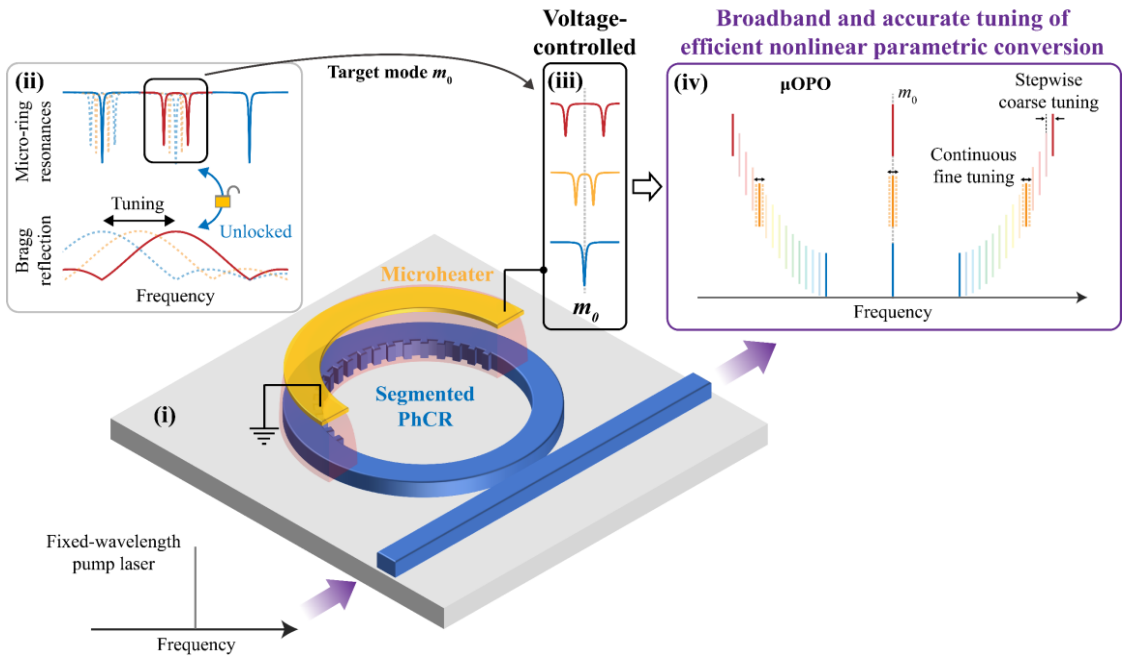

**Fig. 1 Concept of electrically reconfigurable PhCR for high-performance nonlinear convertor.** (i) The electrically reconfigurable PhCRs consists of a segmented PhCR and a microheater positioned on top of the grating region. (ii) The grating distribution angle provides an additional degree of freedom that unlocks the split mode number from half the grating number. Electric tuning of the grating segment can now dynamically aligns or misaligns the Bragg reflection peak with the target MRR resonance. (iii) When different voltage values are applied to the microheater, the target mode splitting is dynamically adjusted. (iv) Our proposed scheme enables high-performance on-chip nonlinear parametric conversion within a single integrated microcavity using a fixed-wavelength pump laser, achieving unprecedented performance with simultaneous broadband output tunability, accurate pump-mode control and high power efficiency.

In contrast to the full-circumference PhCRs, the novel segmented PhCR design introduces an additional degree of freedom, i.e., the grating distribution angle. This innovation physically decouples the fixed relationship between the Bragg relfection frequency $v_B$ and the mirco-ring resonance $v_m$. Similar to the spatially separated coupled micro-rings, the segmented grating allows for differential tuning of the grating region and the entire micro-ring. By applying varying voltages to the microheater, the detuning between Bragg reflection peak and the target resonance can be continuously adjusted. The resulting change in the grating reflectivity at the target resonant mode $m_0$ thereby causes the mode coupling rate $\beta$ to vary from maximum to zero.

The voltage applied to the microheater enables dynamic control of the target mode splitting, inctroducing hitherto unavailable tunability for wavenumber-accurate nonlinear conversion in PhCRs. For instance, when one of the pair of split modes $v_m^{\pm}$ is utilized as the pump mode, the mode shift $\beta$ alters the frequency mismatch in four-wave mixing (FWM) processes[15]:

$$\Delta v_{\text{FWM}} = 2(v_p \pm \beta) - v_s - v_i = \Delta v_{\text{FWM},0} \pm 2\beta \tag{1}$$

$$2m_p = m_s + m_i \tag{2}$$

where $v_p$, $v_s$ and $v_i$ are the frequenies of the unperturbed pump mode $m_p$, signal mode $m_s$ and ilder mode $m_i$, respectively. The frequency mismatch without mode shift $\Delta v_{\text{FWM},0} = 2v_p - v_s - v_i$ is typically non zero due to group velocity dispersion (GVD). However, a voltage-controlled mode shift $\beta$ enables pefect compensation of $\Delta v_{\text{FWM},0}$ for efficient FWM between arbitrary signal and ilder modes that conserve momentum (Eq. 2), while also suppressing unwanted parasitic processes near the pump mode, thereby facilitating high-efficiency μOPO[7,16]. Similarly, the reconfigurable control of $\beta$ can be used to enhance other nonlinear frequency conversion processes requiring perfect phase-matching, such as four-wave mixing Bragg scattering (FWM-BS)[17], third-harmonic generation (THG)[18], and second harmonic generation (SHG), among others. Addtionally, the controllable mode shift supports spontaneous pulse formation[5] and efficient microcomb generation[6,14] in a more robust manner.

**Design principle of segmented PhCRs**

We now investigate the regulatory mechanism of CW-CCW coupling in PhCRs. According to coupled-mode theory, the CW-CCW coupling rate $\beta_m$ for an azimuthal mode $m$ with transverse electric (TE) polorization can be expressed as follows, assuming the grating modulation is a small perturbation[11] (see **Supplementary Section 1**):

$$\beta_m \propto \int_{-\pi}^{\pi} A(\phi) \cos^2(m\phi) \, d\phi \tag{3}$$

where $A(\phi)$ represents the spatial modulation of the width of the micro-ring waveguide as a function of the azimuthal angle $\phi$. The term $\cos^2(m\phi)$ arises from the standing-wave characteristics during coupling. The coupling rate $\beta_m$ denotes the resonant frequency shift of the two standing-wave modes, which represents superposition states including both CW and CCW contributions. To demonstrate the effects of different spatial modulation structures on mode control, we introduce a virtual continuous spatial frequency $n \in \mathbb{R}$. By substituting $\cos(2m\phi) = 2\cos^2(m\phi) - 1$ and $n = 2m$ into Eq. 3, we derive a set of Fourier transforms $A(\phi) \overset{\mathcal{F}}{\Leftrightarrow} S(n)$ as follows:

$$S(n) = \int_{-\pi}^{\pi} A(\phi) \cos(n\phi) \, d\phi \equiv \mathcal{F}(A(\phi)) \tag{4}$$

$$|\beta_m| = |C \cdot S(n)\delta(n - 2m)| \tag{5}$$

where $S(n)$ represents the spectral density in the continuous angular spatial frequency domain, physically corresponding to the continuous reflection spectrum of a non-periodic spatial modulation. The Kronecker delta function $\delta(n)$, which represents the periodicity of the micro-ring, samples from $S(n)$ to yeild physically observable variables, specifically the coupling rate $\beta_m$ at discrete spatial modes ($m \in \mathbb{N}^*$) that resonate within the cavity.

In a full-circumference PhCR with $N_0$ sine-shaped gratings 'teeth', the spatial modulation of the waveguide can be expressed as $A(\phi) = A_0 \cos(N\phi)$, where $A_0$ is the amplitude of the sinusoidal modulation. The effective spatial frequency of the gratings, $N = (2\pi R/\Lambda) \cdot (n_{eff}/n_G) = N_0 \cdot (n_{eff}/n_G)$, is determined by the ratio of the micro-ring's circumference $2\pi R$ to the grating's pitch $\Lambda$, and by the ratio of the micro-ring's effective refractive index $n_{eff}$ to the grating waveguide's effective refractive index $n_G$ (more details in **Supplementary Section 1**). In this case, the entire micro-ring shares the same effective refractive index as the Bragg grating waveguide ($n_{eff} = n_G$), thereby lcoking $N$ to $N_0$.

In the following, we introduce a new degree of freedom—the angular distribution range $\theta$ of the grating segment in the PhCRs (**Fig. 2a**). In segmented PhCRs, the grating waveguide is truncated by an angular window function $G_\theta(\phi_0)$ where $\theta$ and $\phi_0$ represents the width and central position of the window in the azimuthal dimension, respectively. Through Fourier transform, the multiplication of the grating modulation by the truncation window function in the spatial domain corresponds to the convolution of the spectral modulation with the sinc function $S(n)$ in the spatial frequency domain (see **Supplementary Fig. S1**). Segmented PhCRs with varying angular distribution range $\theta$ are fabricated and tested to validate our theoretical model (see **Methods** for details). **Figure 2b** demonstrates the selective mode splitting control in a segmented PhCR, without deterioration of quality factor (intrinsic $Q$ of 1.1 million when $2\beta = 5.56$ GHz). **Figure 2c** shows the measured coupling rate $\beta_m$ (triangles) for these $\theta$-varied devices around the target mode $m = 575$, consistent with our analytic model (solid curves). **Figure 2d** summarizes the $\theta$-dependent measurement of coupling rate $\beta_m$ at the target mode $m = 575$ (blue triangles), the adjacent modes $m = 576$ (orange triangles) and $m = 577$ (yellow diamonds). The solid curves plot the analytic $\beta_m$ for these modes with different $\theta$. As the truncation angle $\theta$ decreases from $2\pi$ to zero, the coupling rate $\beta_m$ linearly decreases at the target mode $m_0 = N_0/2$ (where $N_0$ is even). Additionally, the truncation angle $\theta$ stretches the spectral modulation profile $S(n)$, leading to fluctuations in the coupling rate $\beta_m$ at adjacent, non-targeted modes. A unique scenario occurs at $\theta = \pi$, where $\beta_m$ at all other resonant modes $(m \neq m_0)$ coincides with the zeros of the spectral modulation $S(n)$. Therefore, the half-circumference grating selectively induce a frequency shift at the target mode, similar to previous PhCRs, but not at other modes. It also introduces a new tuning mechanism for dynamic mode conrtol via tunable $N$, controlled by the ratio $n_{eff}/n_G$.

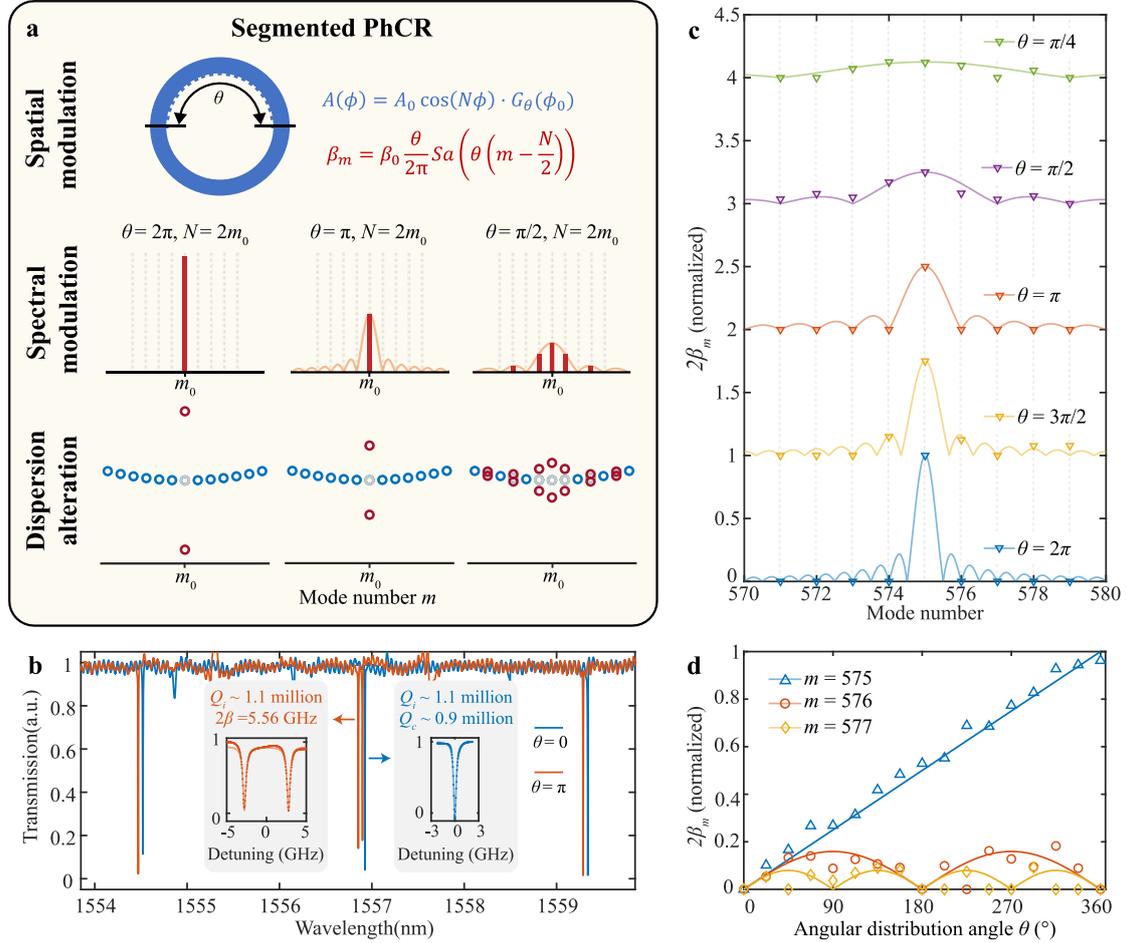

**Fig. 2 Design principle of segmented PhCRs. a** The spatial modulation of the segmented PhCR is the product of a cosine function with modulation frequency $N$ and a window function with an angular distribution range of $\theta$, resulting in a sinc-shaped spectral modulation (orange solid lines) centered at spatial frequency $N/2$. The periodic boundary condition in the micro-ring resonator discretizes the mode frequency (gray dashed lines) and leads to a discrete CW-CCW coupling rate $\beta_m$ (depicted by the red solid lines). This modulation induces a local or broadband alteration (dipicted by red circles) on the original waveguide dispersion profile (dipicted by blue circles) $D_{int}(m) = v_m \pm \beta_m - v_{m_0} - (m - m_0) \cdot FSR$, where $m_0$ is the target azimuthal mode number in the PhCR. **b** Transmission traces of a normal micro-ring ($\theta = 0$, blue) and a segmented PhCR ($\theta = \pi$, orange). Inset: enlarge measured (points) and fitted (lines) results of the target resonance modes in the two devices. **c-d** Normalized mode splitting $2\beta_m$ at different truncation angle $\theta$ with the target mode number set to $m_0 = 575$. The solid lines represent the analytic $\beta_m$, while the marker points indicate the corresponding experimental results.

The discussions above assume a fixed spatial frequency $N = N_0$, typically an even integer. However, in a segmented PhCR, $N$ can be designed to be a non-integer. For instance, the effective spatial frequency can be altered by stretching or compressing the grating waveguide during the design and fabrication, which in turn determines the peak of the spectral modulation $S(n)$ and modifies the coupling rate $\beta_m$. This adjustment can also be effectively achieved by placing a microheater on top of the grating waveguide to change its effective refractive index, as illustrated in **Fig. 1**. Consider a half-circumference segmented PhCR ($\theta = \pi$) with electric tuning applied to the grating waveguide region. The effective spatial frequency $N$ is governed by:

$$N = \left(\frac{2\pi R}{\Lambda}\right) \cdot \left(\frac{n_{eff}}{n_G}\right) = N_0 \cdot \left(\frac{n_G + \Delta n_G \cdot \frac{\theta}{2\pi}}{n_G + \Delta n_G}\right) \approx N_0 \cdot \left(1 - \frac{\Delta n_G}{2n_G}\right) \quad (6)$$

where $n_{eff} = n_G + \Delta n_G \cdot \frac{\theta}{2\pi}$, with $\Delta n_G$ representing the change in effective refractive index in the grating region. In a segmented PhCR with a $\theta < 2\pi$, the effective refractive index of the entire resonator changes less because the accumulated phase shift is averaged over the entire circumference of the microring. Thus, increasing $\Delta n_G$ results in an effective spatial frequency shift $\Delta N = N_0 \cdot \frac{\Delta n_G}{2n_G}$, leading to a translation of the spectral modulation:

$$S(n, \Delta N) = \frac{\beta_0}{2} \text{Sa}\left(\frac{(n-N)\pi}{2}\right) = \frac{\beta_0}{2} \text{Sa}\left(\frac{(n-(N_0 - \Delta N))\pi}{2}\right) \quad (7)$$

$$\beta_m(\Delta N) = S(n)\delta(n - 2m) = \frac{\beta_0}{2} \text{Sa}\left(\frac{(2m-(N_0 - \Delta N))\pi}{2}\right) \quad (8)$$

Where the sampling function $\text{Sa}(n) = \sin(n)/n$ is the fourier transform pair of the window function $G_\theta(\phi_0)$, and $\beta_0$ represents the coupling rate in a full-circumference PhCR with the same spatial modulation amplitude. By applying a voltage to the heater, the split mode number can shift from $m_0$ to $m_0 - 1$ when $\Delta N = 2$ (i.e., a phase modulation of $2\pi$). During this process, the target mode splitting (for $m = m_0$ or $m_0 - 1$) is dynamically tuned between zero and maximum.

### Tuning mechanism of the electrically reconfigurable PhCR

To experimentally demonstrate an electrically reconfigurable PhCR for nonlinear covnersion, we fabricated a segmented PhCR device similar to that shown in Fig. 1, but with an additional heater B located on the unmodulated ring waveguide region, as depicted in **Fig. 3a**. The auxiliary heater B compensates for the thermal drift induced by heater A during mode splitting tuning. The grating segment and heaters are designed to avoide covering the coupling region while maximizing the tuning range of the thermo-optic phase shift. The fabrication process ultilizes our low-temperature deuterated silicon nitride (SiN$_x$) platform, as reported earlier[19,20] (see **Methods**). The resonator features a FSR of 297 GHz, with a waveguide height of 650 nm and a mean width of 2200 nm. The grating has a modulation amplitude of 90 nm, and its spatial frequency is set to $N_0 = 565 \times 2$, enabling selective mode splitting of 7.6 GHz at resonance mode $m_0 = 565$.

The mode splitting and central frequency of the target mode $(m_0 - 1)$ is dynamically controlled by the applied power to heaters A ($P_A$) and B ($P_B$), as shown in **Fig. 3b**. The coefficient $a$ and $b$ denote the resonant frequency tuning factor of the heaters A and B in the unit of GHz/mW. The central frequency shift is affected by both $P_A$ and $P_B$ with tuning factors of $a$ and $b$, while the spatial modulation frequency of the grating is merely determined by $P_A$ with a tuning factor of $2a$. Therefore, the effective spatial modulation frequency shift can be controlled by the detuning $\Delta N$ between Bragg reflection frequency $\nu_B$ and target resonance frequency $\nu_{m0}$, in proportion to $(a \cdot P_A - b \cdot P_B)$.

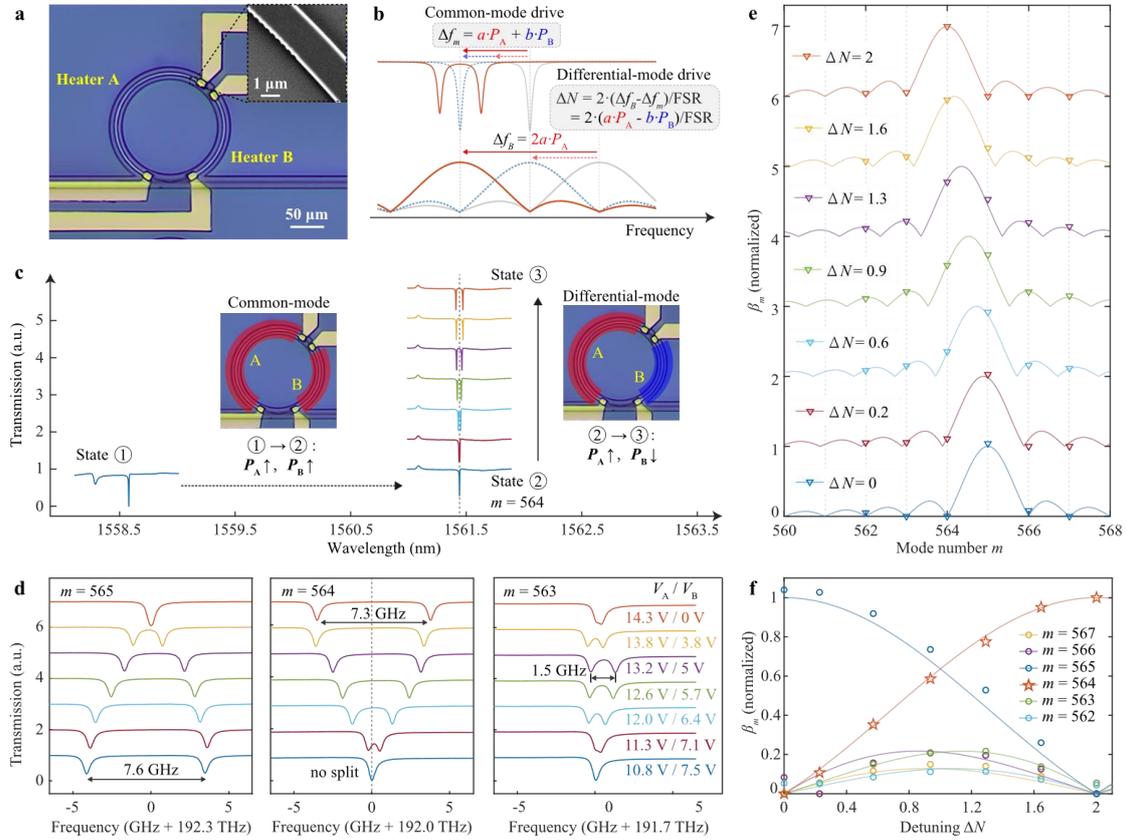

**Fig. 3 Tuning mechanism of the electrically reconfigurable PhCR. a** Microscope image of a fabricated half-circumference segmented PhCR. Inset: an enlarged electron microscope image showing the junction between the grating waveguide and the unmodulated waveguide in the segemented PhCR. **b** Tuning principle of the electrically reconfigurable PhCR. **c-d** Measured transmission traces under different voltages applied to heaters A and B. **e-f** Normalized mode coupling rate $\beta_m$ at different effective modulation frequency detunings $\Delta N = N - 2m_0$ with a target mode number of $m_0 = 565$. The solid lines represent the analytic $\beta_m$, while the marker points indicate the corresponding experimental results. **c** Strategy for dynamic mode splitting control that mitigates thermal drift of all cavity resonances, ultilizing two complementary heaters to stabalize the cavity temperature.

**Figure 3c** illustrates the tuning strategy for independent control of the mode splitting and the central frequency. Inniatially, the PhCR is driven by common-mode voltages (with both $P_A$ and $P_B$ applied) to set an initial central frequency. Subsequently, the PhCR is driven by differential-mode voltages (with $P_A$ increasing as $P_B$ decreases, and vice versa, shown as red and blue, respectively), which keeps the effective refractive index $n_{eff}$ of the entire micro-ring constant while modulating the effective index $n_G$ of the Bragg grating waveguide. **Figure 3d** demonstrates the dynamic control of mode splitting at target resonance mode $m = 564$ (or 565) from 7.3 GHz (zero) to zero (7.6 GHz), thereby introducing conrtollable dispersion alteration for flexible nonlinear conversion. The central frequency remains fixed during the mode splitting control by this complementary heating scheme, thereby eliminating the need for an expensive, widely tunable laser source and enabling applications based on fixed pump frequency operation. Compared to the single-heater method, a disadvantage of this approach is the addtitional static power required to maintain a higher average temperature in the cavity. A potential solution is to use suspended themo-optic phase shifters, which could reduce the electric power consumption by a factor of ten [20].

**Figure 3e** and **3f** illustrate the dependence of $\beta_m$ on the effective detuning $\Delta N$ calculated by $P_A$ and $P_B$ (more details in **Supplemantary Section 2**). When the PhCR is driven by the differential-mode voltages, the spectral modulation center exhibits a gradual red shift toward the resonance mode $m = 564$. The measured mode coupling rate $\beta_m$ of the target resonance mode $m = 564$ increases continually from zero to maximum (orange pentacles) as $\Delta N$ increases from zero to 2, which is consistent with our analytic results (solid lines). In contrast, the coupling rate $\beta_m$ of the other target mode $m = 565$ experiences a gradual decline as $\Delta N$ increases. Although the coupling of counter-porpagating waves at other adjacent modes (e.g., $m = 562, 563, 566, 567$ in **Fig. 3e**) cannot be fully suppressed during the tuning process, it is typically much smaller than the coupling at the target mode and enables tunable nonlinear convertor such as μOPOs.

**Tunable and efficient Kerr μOPO**

We employed a reconfigurable 297-GHz PhCR device with an anomalous GVD of $D_2/2\pi = 7.6$ MHz to demonstrate a tunable Kerr μOPO - a typical nonlinear wavelength convertor that generally requires a widely tunable pump laser due to their limited tunability. Our device operates in an over-coupling regime to enhance the μOPO conversion efficieny, featuring an intrinsic $Q$ of $1.3 \times 10^6$ and a loaded $Q$ of $0.43 \times 10^6$. We amplified the pump laser to 195 mW (on-chip) at the target mode $m_0 = 574$ to generate signal and ilder emissions. The generated signal and idler modes are determined by the μOPO frequency matching, which is quantified with the integrated dispersion,

$$D_{int}(m) = v_m \pm \beta_m - v_{m_0} - (m - m_0) \cdot \text{FSR} \qquad (9)$$

where $m_0$ is the target pump mode number in the PhCR ($m_0 = 574$ in our case). **Figure 4a** illustrates the tuning principle of our µOPO. Appropriate Kerr frequency shifts are included in the hot-cavity resonator mode frequencies $v_m$, which is twice as large for $m \neq m_0$ than the pump mode $m = m_0$. By controlling the CW-CCW coupling-induced frequency shift $\beta$ of the target pump mode via the voltages applied to the heaters, the phase-matching modes, as well as the generated signal/idler wavelengths in the µOPO, can be dynamically tuned at a discrete step of 1 FSR. The numerical simulation are presented in **Supplementary Section 3**, where the effect of adjacent mode splittings during the tuning process are investigated. With increased electric power applied to heater A, the mode splitting of $2\beta$ is dynamically reconfigured from 0 to 5 GHz (corresponding a dispersion alteration $\beta$ varied from 0 to 2.5 THz) at a slope efficiency of ~0.046 GHz/mW, while the frequency of the resonance exhibits a thermal drift in proportion to the applied power (**Fig. 4b**).

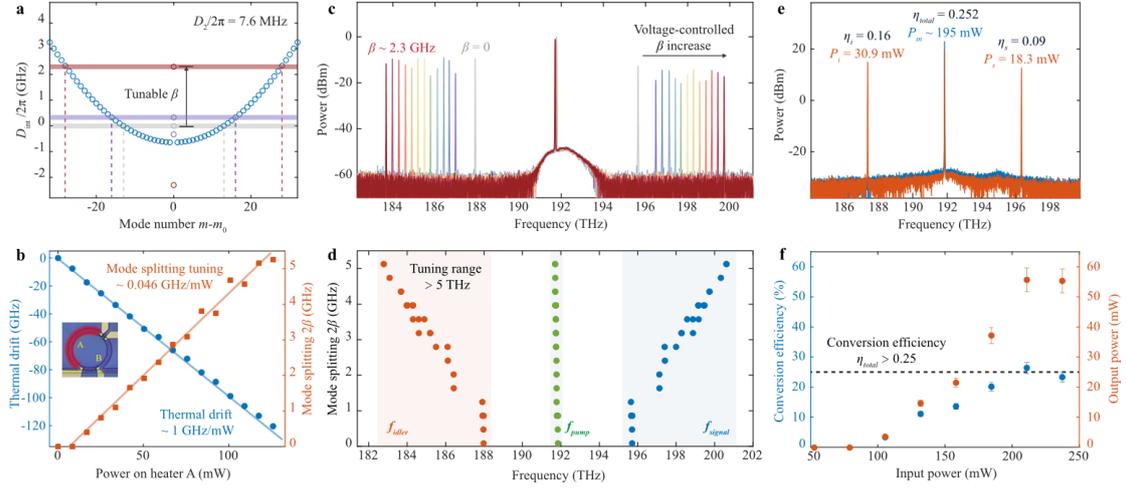

**Fig. 4 Tunable Kerr µOPO enabled by voltage-controlled mode shift in PhCR. a** Tuning principle of a tunable Kerr µOPO enabled by a reconfigurable mode shift in the integrated dispersion profile. The circles represent the pumped modes ($m = 574$) at different $\beta$ values: $\beta = 0$ GHz (gray), $\beta = 0.33$ GHz (purple), $\beta = 2.3$ GHz (red) and non-pumped modes exhibiting additional Kerr phase shift (blue). The solid lines depict the phase-matching condition (bold) and the particitating modes (dashed) in the PhCR-based µOPO at various $\beta$ levels, indicated by the corresponding colors. **b** Measured mode splitting $2\beta$ (orange squares) and thermal drift (blue points) with varying electric power on Heater A, with the data fitted to lines. **c** Measured optical spectra of a single µOPO operating at different $\beta$ values, controlled by varying the voltages applied to heater A. **d** Measured frequencies of pump (green), idler (orange) and signal (blue) light at different $\beta$ values. **e** Measured optical spectra of the µOPO at the highest power conversion efficiency of 0.252 (orange), alongside the unconverted pump light at an off-resonance frequency (blue). **f** Measured total power conversion efficiency $\eta_{total}$ and summed output power $P_{out} = P_s + P_i$ at varying input power.

**Figure 4c** shows the experimental measurement of the output spectra obtained in a single PhCR with ~195 mW (on-chip) pump light near 191.7 THz, depicted by different colors for different operating $\beta$ values. By electrically adjusting $\beta$ from 0 to 2.5 GHz, we successfully achieve efficient μOPO with tuability of signal/idler frequencies over a 5 THz range (**Fig. 4d**). This approach eleminates the need for an expensive widely tunable laser. The maximum tuning range is primarly constrained by the cavity's GVD and the available $\beta$, which can be effectivetly improved by adopting a cavity with smaller GVD near the zero-dispersion region and increasing the modulation amplitude of the grating. By superimposing multi-period ($N_0$) modulation and implementing coherent synthesis in segmented PhCRs, it is anticipated that the dispersion alteration $\beta$ can be further expanded without compromising the $Q$ factor[21,22].

Then, we investigate the power conversion efficiency $\eta = P_{out}/P_{in}$, which is defined as the ratio of the output power $P_{out} = P_s + P_i$ (the sum of signal and idler power) to the input power $P_{in}$. We measured the output power and conversion efficiency of our μOPO device at different input power, achieving a maximum on-chip conversion efficiency of >25% at ~200 mW (on-chip) input power at frequency near 191.7 THz (**Fig. 4f**). **Figure 4e** presents the output μOPO spectrum obtained in one of the most efficient state with $\beta$ of ~1 GHz. The signal and idler light power are 18.3 mW and 30.9 mW respectively, corresponding a total power conversion efficiency of 0.252. We note that only forward propagating light at the through port is included for energy efficiency calculation here. We anticipate that a light reflector at the throught port for pump recycling[23] will help to improve conversion efficiency and lower threshold power when the μOPO is operated at a large $\beta$.

While the above experiment demonstrates step-wise precise selection of signal-idler modes with fixed pump mode, we next investigate gap-free, broadband tuning of the generated signal and idler light to achieve continuous spectral coverage in the μOPOs. The tuning strategy is comprised of discrete broadband tuning under differential mode operation (as shown above) and continuous fine tuning under common mode operation.

In the differential mode operation, the power on heater A ($P_A$) and the power on heater B ($P_B$) change inversely and compensate each other in minimizing total thermal drift while adjusting the effective detuning between the grating modulation frequency and the resonance frequency. **Figure 5a** shows the thermal drift-free mode splitting control of the target pump mode, where the effective detuning is derived from $P_A$ and $P_B$ (see **Supplemantary Section 2**). The differential mode operation enables broadband frequency switching of the signal and idler modes with a fixed pump laser. **Figure 5b** demonstrate signal (idler) mode number switching, spanning 13-FSR range from 587 (561) to 600 (548) with pump frequency fixed at 191.732 THz ($m = 574$).

In the common mode operation, $P_A$ and $P_B$ change in the same direction in a specific proportion, linearly shifting the resonant frequencies of all the resonator modes without affecting the detuning between the grating modulation frequency and the resonance frequency. As shown in **Figure 5c**, we emulates the common-mode heating by adjusting the overall device temperature from 10°C to 60°C via the thermoelectric cooler (TEC). The central frequencies of all resonances are globally tuned with a thermal tuning efficiency of ~2 GHz/°C. Notably, the target mode coupling rate $\beta$ exhibits excellent thermal stability, with a variation of less than 0.04 GHz within a frequency tuning range of >100 GHz. This stability in response to overal temperature fluctuations is crucial for high-power nonlinear conversion processes, which are often affected by strong thermal effects. Moreover, it enables independent tuning of the pump resonance frequency without affecting the dispersion alteration $\beta$, thereby translating the continuous frequency shift to signal and idler lights. **Figure 5d** demonstrates the TEC-based fine tuning of the signal and idler lights in the μOPO, showing a tuning efficiency of ~2 GHz/°C. In the future, the common-mode heating can be integrated into the two microheaters by improving the heating efficiency of the thermo-optic phase shifters.

**Figure 5e** illustrates the tuning strategy to achieve continuous spectral coverage of the overal μOPO tuning range. The output signal and idler frequencies are controlled by reconfigurable dispersion alteration $\beta$ and continuously tunable pump frequency $f_p$:

$$f_{s/i} = f_p \pm n(\beta) \times \text{FSR} \qquad (10)$$

where $n$ is the sideband-to-pump mode spacing, determined by the dispersion alteration $\beta$. The reconfigurable control of $\beta$ enables μOPO state switching among different sideband-to-pump spacings, facilitating broadband tuning of the output frequency in the step size of 1 FSR. Additionally, the output frequency can be continually tuned by simultaneously controlling the global resonance frequency $f_m$ and the pump laser frequency $f_p$, without affecting the laser-resonance detuning. In this scheme, only a narrowband tunable (slightly larger than 1 FSR) laser source is required to achieve gap-free, broadband tuning of the μOPO. It is noteworthy that by hybridly integrating a laser diode with the μOPO, the pump laser frequency can be locked to the micro-ring by means of self-injection locking[10], thereby can follow the fine tuning without additional control mechanism.

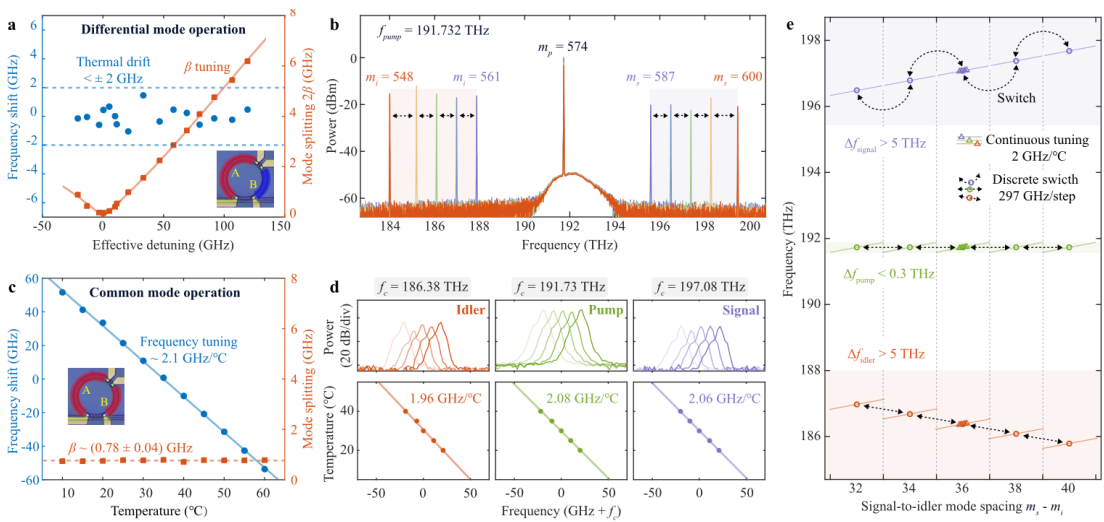

**Fig. 5 Gap-free, broadband tuning strategy for full spectral coverage. a** Measured frequency

shift and mode splitting of the target pump resonance at different effective detuning, controlled by electric power $P_A$ and $P_B$. **b** Measured optical spectra of a single µOPO operating at varying $\beta$ values using a frequency-fixed pump laser. The µOPO outputs are switchable in the step size of single-FSR. For clarity, the selected optical spectra with signal-to-idler mode spacings of 26 (purple), 32 (blue), 38 (green), 44 (yellow) and 52 (orange) are presented. **c** Measured frequency shift and mode splitting of the target pump resonance at different overall device temperatures controlled by the TEC. **d** Measured optical spectra (top) and frequency shift (bottom) of the pump (green), idler (orange) and signal (purple) lights at different TEC temperatures. **e** Schematic illustration of gap-free broadband tuning strategy for full spectral coverage, with dynamic control of both dispersion alteration $\beta$ and pump resonance frequency $f_p$. The circles represent experimental results for the pump (green), idler (orange) and signal (purple) lights with varying signal-to-idler mode spacings. The triangles denote measured frequencies of the pump, idler and signal lights within the 36-FSR signal-to-idler spacing region as the TEC temerature varied from 20°C to 40°C. The solid lines ideally indicate contunous tuning traces of the the pump, idler and signal frequencies, while the dashed arrows depict the mode spacing switching processes.

**Discussion**

By dynamically controlling the target mode splitting in a segmented PhCR, we have developed a high-performance µOPO that offers both a broad signal/idler tuning range (exceeding 5 THz) and high energy conversion efficiency (surpassing 25%). Furthermore, the implementation of complementary heaters facilitates wavelength-fixed pump operation, enabling exceptional output tunability without the reliance on an expensive, widely tunable laser source.

**Table 1** summarizes the state-of-art Kerr µOPO demonstrations. Among the reported Kerr µOPOs, our work showcases unprecedented capability to achieve both high conversion efficiency and extensive frequency tunability. This achievement is attributed to the reconfigurable dispersion alteration $\beta$, which allows for perfect phase-matching for arbitrary target signal and idler modes within the tuning range. The spectral coverage and output tuning range demonstrated in this work are respectively modest compared to other fixed µOPO due to suboptimal dispersion engineering (rather than the reconfigurable PhCR structure itself) and can be further improved. We anticipated that further optimization of the global dispersion will enable a widely tunable µOPO covering the mid-infrared[24] or visible light[25] region of interest.

In addition to broadband tuning in a discrete manner, the ability to achieve continuous tuning is crucial, particularlly in quantum systems with specific transition wavelengths. Our work presents a novel gap-free solution for broadband spectral coverage that combines reconfigurable dispersion alteration for discrete broadband tuning with a global frequency offset for continuous fine tuning, all within a single µOPO utilizing two complementary microheaters. To achieve continuous spectral coverage, a continuous frequency tuning range of grater than one FSR is required. Some efficient phase-shifting technolygies, such as using isolation trenches and suspended waveguides[20], may be beneficial in this context.

Unlike previous approaches to broadband dispersion tuning using coupled micro-rings, the accurate mode selectivity of our dispersion-reconfigurable device efficiently surpresses unwanted nonlinear parasitic processes near the pump mode, thereby enabling power-efficient μOPO. Additionally, the proposed electrically reconfigurable PhCR structure is compatible with advanced efficiency enhancement schemes, such as utilizing bound states in the continuum (BICs)[26] or implementing a pump energy recycling mechanism[23]. Furthermore, the grating segment embedded within the micro-ring resonator provides a wavelength-accurate target mode for wavelength-fixed pump laser, facilitating the foundry manufacturing of chip-scale μOPO integrated with the pump laser source.

More generally, electrically reconfigurable PhCRs offers exceptional capability of accurate, dynamic dispersion control, providing a versatile platform for fundamental researchs such as nonlinear conversion and soliton dynamics. The reflectivity control during the dispersion tuning process also brings new insight into self-injection locking technologies in generating soliton pulse[10] and harmonic-waves[27]. Notably, while the PhCR demonstrated in this work utilizes a single spatial modulation frequency, the introduction of multiple-color grating modulation could open up novel applications, such as real-time tailoring of microcomb spectra[21,22].

In conclusion, we have theoretically proposed and experimentally demonstrated an electrically reconfigurable PhCR that enables dynamic target mode control, facilitating flexible nonlinear wavelength conversion. A high-performance μOPO with both high frequency tunability and high conversion efficiency is successfully demonstrated, without the need for a widely tunable laser source. Future optimization of resonator dispersion, available mode splitting and pump energy recycling could further expand the spectral coverage and enhance conversion efficiency. Looking ahead, multi-target-mode tuning using multi-period segmented PhCRs could enable advanced applications, such as multi-mode nonlinear conversion and dynamic comb spectrum tailoring. Importantly, we anticipate that our dynamic dispersion control approach can be extended to resonant $\chi^{(2)}$-nonlinear systems, in addition to the $\chi^{(3)}$ systems explored here. The devices and methods introduced in this work are expected to be valuable for future chip-based applications that require efficient and reconfigurable nonlinear photonics.

**Table 1. Comparison of the demonstrated Kerr μOPOs.**

| Ref. | Scheme | $\lambda_{input}$ tuning range | $\lambda_{output}$ tuning range | Tuning ratio ($\frac{\delta\lambda_{out}}{\delta\lambda_{in}}$) | Step size | Thermally continuous tuning | Specral coverage | Dispersion regime | $P_{out}$ | $P_{out}/P_{in}$ |
|---|---|---|---|---|---|---|---|---|---|---|
| [28] | Single MRR | ≈5 THz | >130 THz | 26:1 | Multi-FSR | Adjacent mode hopping | 273.7 THz | Normal $D_2 \approx 0$ | <0.01 mW | <0.1 % |
| [29] | Single MRR | / | / | / | Multi-FSR | / | ≈346 THz | Normal $D_2 \approx 0$ | <2 mW | ≈1% |
| [30,31] | Single MRR | ≈5 | ≈95 | 19:1 | Multi- | 0.3 THz | >150 THz | Strong | >18 | 29% |

| | | THz | THz | | FSR | ≈0.3 FSR | | normal | mW | |
|---|---|---|---|---|---|---|---|---|---|---|
| | (hybrid-mode OPO) | | | | | | | | | |
| [7] | PhCR | Fixed | Fixed | 0 | / | 50 GHz ≈0.05 FSR | >100 THz | Arbitrary | ≈4 mW | >10% |
| [23] | PhCR+ reflector | Fixed | Fixed | 0 | / | / | ≈15 THz | Arbitrary | 40 mW | 41% |
| [8,32] | PhCR | Fixed | Fixed | 0 | / | 0.8 THz ≈0.9 FSR | ≈120 THz | Normal | >2 mW | 12.5% |
| [26] | BIC | Fixed | Fixed | 0 | / | / | ≈7 THz | Anomalous | ≈136 mW | 68% |
| [12] | Coupled rings | ≈20 GHz | ≈20 THz | 1000:1 | Single-FSR | / | ≈45 THz | Normal-anomalous | / | <0.1% |
| This work | Segmented PhCR | Fixed | > 5 THz | +∞ | Single-FSR | 82 GHz ≈0.3 FSR | ≈19 THz | Arbitrary | ≈50 mW | >25% |

**Methods**

The SiN$_x$ film, with a thickness ranging from 650 nm to 850 nm, is deposited on a silicon wafer with a 3 μm thick thermally-grown silica (SiO$_2$) layer at 270 °C, using deuterated silane (SiD$_4$) and pure nitrogen (N$_2$) as the source gases. This layer is deposited in single continuous run utilizing the inductively coupled plasma chemical vapor deposition (ICP-CVD) without high-temperature annealing or chemical-mechanical polishing (CMP). The devices in this study are fabricated exclusively through subtractive processing. The PhCR patterns are defined via electron beam lithography (EBL) on an 800 nm thick AR-P 6200 resist layer and subsequently transferred to the SiN$_x$ film by reactive ion etching (RIE) using CHF$_3$ and O$_2$ gases.

After etching, the waveguides are stripped of photoresist and cladded with a 3 μm ICP-CVD silica layer, deposited using silane ($SiH_4$) as the source gas. The top cladding is planarized and thinned to $\approx 1.75$ μm through spin-coating of AZ 2035 photoresist followed by etching back. Ni-Cr alloy heaters, along with Au wires and electrodes, are patterned using EBL and then deposited by electron beam evaporation (EBE). An inverse taper design is employed to increase mode size of light and reduce chip-to-fiber coupling losses.

The characterization setup for both linear and nonlinear measurements is shown in **Supplementary Fig. S4**. The devices under test are mounted on a thermally stable platform, which is temperature-controlled using a TEC. Input and output light are coupled into and out of the chip via lensed fibers. In the linear measurements, a power meter records the transmission traces as the wavelength of the input light is swept using a widely tunable external-cavity diode laser. The coupling rate $\beta$ and central frequency $f_m$ are determined by fitting the resonance data to either a Lorentzian line shape model or a resonance doublet model[33]. For the nonlinear measurements, the pump from the external-cavity diode laser is amplified by an erbium-doped fiber amplifier (EDFA). To suppress amplified spontaneous emission (ASE) of the amplifier, a tunable narrow-band filter is placed after the amplifier. The generated light is analyzed using an optical spectrum analyzer (OSA). Additionally, a photodiode connected to an oscilloscope is used to measure the transmission traces. In both measurements, the PhCR's heater elements are powered by a voltage source, which is critical for tuning the effective spatial modulation frequency detuning $\Delta N$ and the output signal/idler frequencies of the μOPO.

The conversion efficiency is characterized by measuring the output spectrum from the bus waveguide using an OSA. First, we measure the output when the μOPO is operated, summing the powers of the signal and idler lights $(P_s + P_i)$. Then, we turn off the μOPO by detuning the laser frequency completely out of resonance and perform a second measurement on the OSA, recording the power of the unconverted pump light ($P_{in}$). The conversion efficiency is calculated as $(P_s + P_i)/P_{in}$. By measuring the loss between the lensed fibers and our chips, we infer the on-chip input and output power from the off-chip power measured by a power meter, with a uncertainty of 0.3 dB. The power in the spectra of **Fig. 4e** represent the on-chip power, which is calculated from the off-chip spectra measured by the OSA.

**Data Availability**

The data that support the findings of this study are available on request from the corresponding author.

**Code Availability**

The simulation codes used in this study are available on request from the corresponding author.

**Acknowledgements**

This work was supported by National Natural Science Foundation of China (NSFC) (62475291, 62335019).

**Author contributions**

J.L. contributed in the conception, design and fabrication, and performed the optical measurement and theoretical analysis. J.Z. assisted with device fabrication. Y.Z. and S.Y. contributed to the theoretical understanding and supervised the findings of this work. All authors provided feedback and helped shape the research, analysis and manuscript.

**Competing Interest**

The authors declare no competing interests.

# Supplements

# Broadband and Accurate Electric Tuning of On-Chip Efficient Nonlinear Parametric Conversion

*Jiaqi Li[1], Yanfeng Zhang[1,\*], Jinjie Zeng[1], and Siyuan Yu[1,\*]*

[1]State Key Laboratory of Optoelectronic Materials and Technologies, School of Electronics and Information Technology, Sun Yat-Sen University, Guangzhou 510006, China.

*Corresponding author: zhangyf33@mail.sysu.edu.cn; yusy@mail.sysu.edu.cn*

**Section 1. Theoretical analysis of segmented PhCRs**

When the mirco-ring resonant frequency $v_m$ coincides with the Bragg relfection frequency $v_B$, strong coupling between clockwise (CW) and counterclockwise (CCW) propagating modes occurs, resulting in the splitting of the original mode $v_m$ into two standing-wave modes $v_m^\pm = v_m \pm \beta_m$. This splitting corresponds to a frequency shift equal to the CW-CCW coupling rate $\beta_m$, which can be expressed as follows[1]:

$$\beta_m = \frac{v_m}{2} \frac{\int dS \cdot A(\phi) \left[ (\varepsilon_1 - \varepsilon_2)|\boldsymbol{E}_\parallel|^2 + \left(\frac{1}{\varepsilon_2} - \frac{1}{\varepsilon_1}\right)|\boldsymbol{D}_\perp|^2 \right]}{\int dV \cdot \varepsilon \left(|\boldsymbol{E}_\parallel|^2 + |\boldsymbol{E}_\perp|^2\right)} \quad (S1)$$

where $v_m$ represents the frequency of resonant modes, and $A(\phi)$ corresponds to the grating modulation of the micro-ring's inner radius, which is a function of the azimuth angle $\phi$. The parameters $\varepsilon_1$ and $\varepsilon_2$ denote the dielectric constants of the waveguide and cladding materials, respectively. The electric field components $E_\parallel$ and $E_\perp$ represent the fields parallel and perpendicular to the micro-ring's sidewall. For simplicity, we focus exclusively on the TE mode in our theoretical analysis and experimental verification, where the electric displacement vector $\boldsymbol{D}$ takes the form of a standing wave: $\boldsymbol{D}(r,\phi,z) = \boldsymbol{D}(r,z) \cdot \cos(m\phi)$. Here, $r$ and $z$ denote the radial and vertical coordinate, respectively, and $m$ represents the mode number. By simplifying Eq. (S1), we obtain:

$$\beta_m = \frac{v_m}{2} \frac{\int dS \cdot A(\phi) \left(\frac{1}{\varepsilon_2} - \frac{1}{\varepsilon_1}\right) |\boldsymbol{D}(r,z)|^2 \cos^2(m\phi)}{\int dV \cdot \varepsilon |\boldsymbol{E}(r,z)|^2 \cos^2(m\phi)} = C \int_{-\pi}^{\pi} A(\phi) \cos^2(m\phi) \, d\phi \quad (S2)$$

where $C = \frac{v_m}{2} \frac{\int dz \cdot \left(\frac{1}{\varepsilon_2} - \frac{1}{\varepsilon_1}\right)|\boldsymbol{D}(r,z)|^2}{\int dV \cdot \varepsilon |\boldsymbol{E}(r,z)|^2 \cos^2(m\phi)}$. This leads to Eq. 3 in the main text. From Eq. S2, we deduce that $\beta_m$ can be controlled by the modulation amplitude and azimuthal distribution of the Bragg gratings.

Consider a full-circumference photonic crystal micro-ring resonator (PhCR) with $A(\phi) = A_0 \cos(N_0 \phi)$ where $N_0$ is an even integer. By ultilizing the periodicity of the standing-wave term $\cos^2(m\phi)$, we can selectively introduce mode splitting

$$\beta_m = C \cdot A_0 \int_{-\pi}^{\pi} \cos(N_0\phi) \cos^2(m\phi) \, d\phi = \begin{cases} \beta_0, & m = m_0 \\ 0, & m \neq m_0 \end{cases} \quad (S3)$$

where $\beta_0 = CA_0\pi/2$. Since $N_0$ is defined by the ratio of the micro-ring's circumference $2\pi R$ to the grating's pitch $\Lambda$, the splitting mode number is fixed by the physical structure of the full-circumference PhCR.

In segmented PhCRs, the grating waveguide is truncated by an angular window function $G_\theta(\phi_0)$, where $\theta$ and $\phi_0$ represents the width and central position of the window in the azimuthal dimension, respectively. Since the angular position $\phi_0$ affects only the phase characteristics rather than the amplitude of the coupling rate $\beta_m$, we set $\phi_0$ to zero for simplicity in our analysis. Through Fourier transformation, the multiplication of the grating modulation by the truncation window function in the spatial domain corresponds to the convolution of the spectral modulation with the sinc function $S(n)$ in the spatial frequency domain (**Fig. S1**).

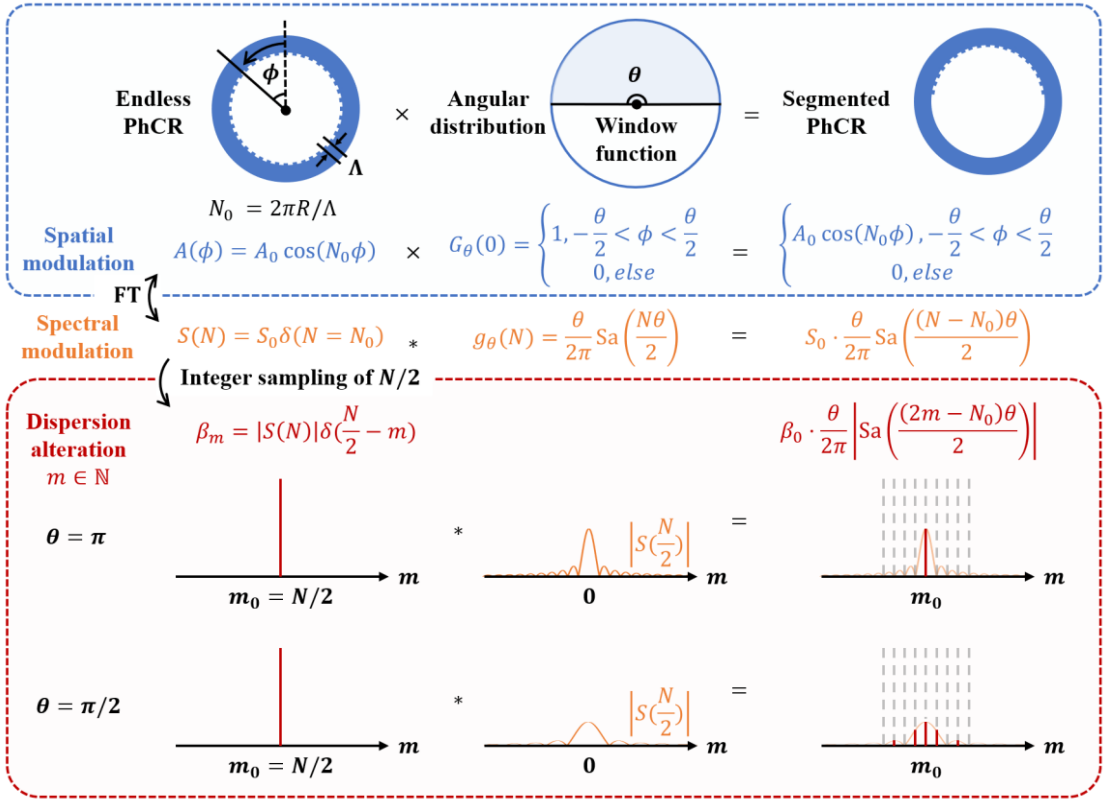

**Fig. S1** Mathematic illustration of segmented PhCRs.

The discussions above assume a fixed spatial frequency $N = N_0$, typically an even integer. However, in a segmented PhCR, $N$ can be designed to be a non-integer. Consider a PhCR with a tunable grating segment (**Fig. S2**), the effective refractive index $n_B$ of the grating waveguide does not always equal to the effective index $n_0$ of the

normal waveguide. The effective refractive index of the entire micro-ring can be expressed as follows:

$$n_{eff} = n_B \cdot \frac{\theta}{2\pi} + n_0 \cdot \left(1 - \frac{\theta}{2\pi}\right) \tag{S4}$$

Furthermore, the uniform standing-wave form of the electric field should be modified by the local effective refractive index $n$ in either the grating region or the normal region:

$$\boldsymbol{E}(r,\phi,z) = \boldsymbol{E}(r,z) \cdot \cos\left(m \cdot \frac{n}{n_{eff}} \cdot \phi\right) \tag{S5}$$

By replacing the physical azimuth angle $\phi$ with the effective optical path $\phi' = (n/n_{eff}) \cdot \phi$, we can map the spatial modulation onto an equidistance grid with an optical phase spacing of $2\pi$:

$$A(\phi') = \begin{cases} A_0 \cos\left(\frac{n_{eff}}{n_B} N_0 \phi'\right), & -\frac{\theta}{2} \cdot \frac{n_B}{n_{eff}} < \phi' < \frac{\theta}{2} \cdot \frac{n_B}{n_{eff}} \\ 0, & else \end{cases} \tag{S6}$$

The effective spatial frequency $N$ is defined as follow to account for the optical path distortion caused by the uneven effective refractive index across the entire micro-ring:

$$N = \frac{n_{eff}}{n_B} N_0 \tag{S7}$$

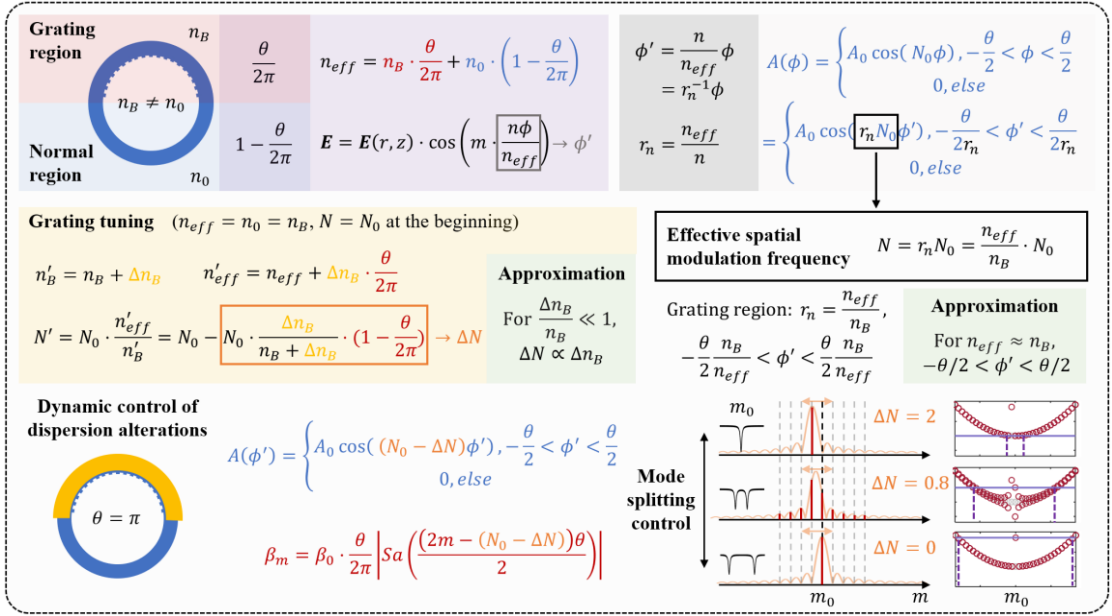

**Fig. S2** Mathematic discription of reconfigurable PhCRs.

In a reconfigurable segmented PhCR with a microheater, local heating results in an increment $\Delta n_G$ of the effective refractive index in the grating region, which corresponds to an increment of $\Delta n_{eff} = \Delta n_B \cdot \theta/2\pi$ averaged over the entire micro-ring. Assuming $n_{eff} = n_0 = n_B$ and $N = N_0$ before tuning, the effective spatial frequency can be expressed as follows:

$$N = N_0 \cdot \left( \frac{n_G + \Delta n_G \cdot \frac{\theta}{2\pi}}{n_G + \Delta n_G} \right) \approx N_0 - N_0 \cdot \frac{\Delta n_G}{n_G} \cdot \left(1 - \frac{\theta}{2\pi}\right) = N_0 - \Delta N \quad (S8)$$

where the effective spatial frequency detuning $\Delta N \propto \Delta n_G$ under a small perturbation approximation ($\Delta n_B \ll n_B$). Furthermore, we approximate the effective grating distribution angle $\theta \cdot (n_B/n_{eff}) \approx \theta$, which is valid for small variances of $N$ (i.e., $\Delta N/N < 0.01$), as shown in **Fig. S3**. Thus, we introduce the effective spatial frequency $N$ while retaining the notation of $\phi$ in the main text. Consequently, we derive the analytic expression for $\beta_m$:

$$\beta_m(\Delta N) = C \cdot A_0 \int_{-\frac{\theta}{2}}^{\frac{\theta}{2}} \cos\big((N_0 - \Delta N)\phi\big) \cos^2(m\phi) \, d\phi \quad (S9)$$

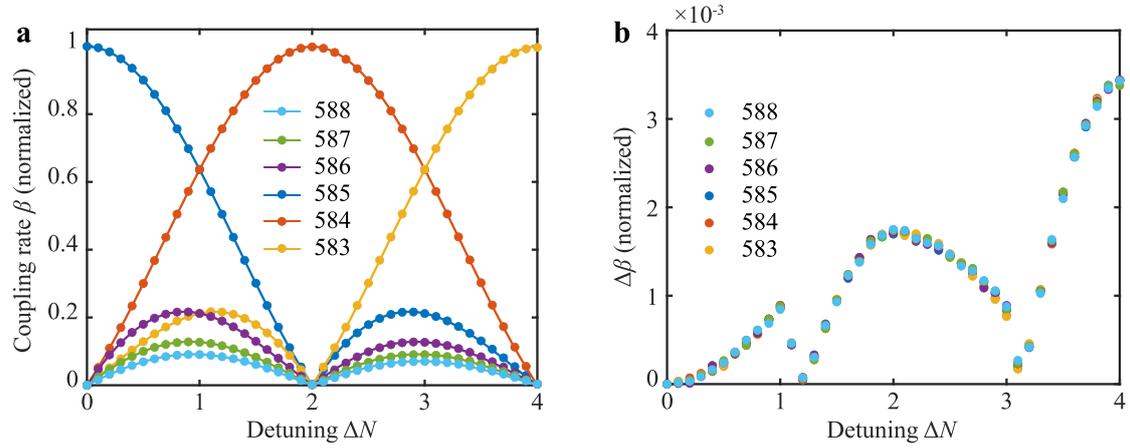

**Fig. S3 The effect of distribution angle correction on $\beta_m$.** **a** Analytic calculation of $\beta_m$ for resonant modes ($m = 583, 584, 585, 586, 587, 588$) at different detunings $\Delta N$, shown with (solid points) and without (solid lines) correction of the distribution angle $\theta$. **b** The difference between the $\beta_m$ values with and without correction, normalized by the maximum $\beta$, is less than 0.004.

## Section 2. Design and characterization of reconfigurable PhCRs

Experimentally, we design the modulation amplitude of the ring width using a square-wave function:

$$A(\phi) = \begin{cases} \dfrac{L}{2}, \dfrac{(2n-0.5)\pi}{N_0} \leq \phi < \dfrac{(2n+0.5)\pi}{N_0}, n = -\left[\dfrac{\theta N_0}{2\pi}\right], \dots, \left[\dfrac{\theta N_0}{4\pi}\right] \\ -\dfrac{L}{2}, \dfrac{(2n-1.5)\pi}{N_0} \leq \phi < \dfrac{(2n-0.5)\pi}{N_0}, n = -\left[\dfrac{\theta N_0}{2\pi}\right], \dots, \left[\dfrac{\theta N_0}{4\pi}\right]+1 \\ 0, \quad else \end{cases} \quad (S10)$$

where $L$ represents the radial length of the gratings, $N_0 = 2\pi R/\Lambda$ is the modulation frequency of the gratings, with $R$ denoting the mean radius of the micro-ring and $\Lambda$ representing the pitch of the gratings. The grating distribution angle is denoted by $\theta$. The duty circle of the grating is set to 0.5 to achieve maximum mode splitting. It is worth noting that the square-shaped grating exhibits the same mode coupling features as the sine-shaped grtaing discussed in the previous section, with a linear amplification factor for the $\beta_m$ amplitude across all resonant modes.

To independently control both the coupling rate $\beta$ and the central frequency $f_m$ of the target resonance, we implement two complementary heaters in our reconfigurable PhCR devices (see Fig. 3a in the main text). According to Eq. S9, the target mode coupling rate $\beta$ is determined by the effective spatial frequency detuning $\Delta N$. Initially, we consider $f_B = f_m = f_0$ and $n_{eff} = n_B = n_0$. The powers applied to heater A ($P_A$) and heater B ($P_B$) dynamically adjust the effective refractive index in the heater regions (with $\theta_A = \pi$ and $\theta_B = \pi/2$) to control $\Delta N$ and $\Delta f_m$ (see Fig. 3b in the main text):

$$\Delta n_{eff} = \Delta n_B \cdot \dfrac{\theta_A}{2\pi} + \Delta n_0 \cdot \left(\dfrac{\theta_B}{2\pi}\right) = \dfrac{\Delta n_B}{2} + \dfrac{\Delta n_0}{4} = a_0 \cdot P_A + b_0 \cdot P_B \quad (S11)$$

$$\Delta n_B = 2a_0 \cdot P_A \quad (S12)$$

$$\Delta f_m = \dfrac{m \cdot c}{2\pi R} \cdot \Delta(n_{eff}^{-1}) = \dfrac{f_0}{n_0} \cdot \Delta n_{eff} = a \cdot P_A + b \cdot P_B \quad (S13)$$

$$\Delta f_B = \dfrac{c}{2\Lambda} \cdot \Delta(n_B^{-1}) = \dfrac{f_0}{n_0} \cdot \Delta n_B = 2a \cdot P_A \quad (S14)$$

$$\Delta N = N - N_0 = \dfrac{2(\Delta f_B - \Delta f_m)}{\text{FSR}} = \dfrac{2(a \cdot P_A - b \cdot P_B)}{\text{FSR}} \quad (S15)$$

where $a_0$ and $b_0$ are the effective refractive index tuning factors for heaters A and B. The parameters $a$ and $b$ are the corresponding central frequency tuning factors for heaters A and B, which can be experimentally determined and used to calculate the effective spatial frequency detuning $\Delta N$.

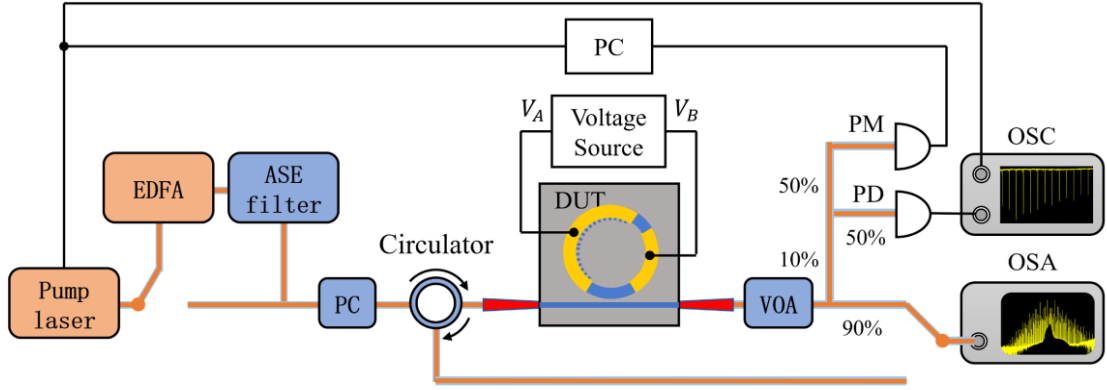

**Fig. S4** Characterization setup. PC: personal computer; EDFA: erbium doped fiber amplifier; ASE filter: amplified spontaneous emission filter; PC: polarization controller; DUT: device under test; PM: power meter; PD: photodiode; OSC: oscilloscope; OSA: optical spectral analyzer.

The characterization setup for both linear and nonlinear measurements are shown in **Fig. S4**. With zero power on the heaters, the Bragg reflection peak aligns with the target resonance and the detuning $\Delta N$ is near zero. A slight deviation from zero, along with non-zero adjacent mode splittings, can be compensated by applying a small offset to $P_A$ for $\Delta N < 0$ (or an offset to $P_B$ for $\Delta N > 0$). We propose two methods to obtain the detuning $\Delta N$ (**Fig. S5**).

The first method requires a tuning range for $\Delta N$ larger than 2, allowing the split mode number to shift from $m_0$ to $m_0 - 1$. The perfect alignment condition is determined by achieving a minimum adjacent mode splitting. Subsequently, the detuing $\Delta N$ along the tuning path can be calculated by the ratio of the actual power difference to the standard power difference for $\Delta N = 2$. The measured detuning $\Delta N$ presented in the main text Fig. 3f is based on this method and allows good consistency with the analytic results.

However, for devices with a smaller tuning range ($\Delta N < 2$), such as the PhCR used for the OPO demonstration in the main text (Fig. 4 and Fig. 5), a standard power difference cannot be measured. To address this, we calibrate the frequency tuning factors $a$ and $b$ by calculating the ratio of the central frequency shift to the applied power at heater A and B (Fig. 3). This allows us to calculate the detuning $\Delta N$ for arbitray combinations of $P_A$ and $P_B$. The second method is employed for effective detuning estimation in the main text (Fig. 4c-d and Fig. 5a), using calibrated tuning factors of $a = 0.95 \text{ GHz/mW}$ and $b = 1.03 \text{ GHz/mW}$.

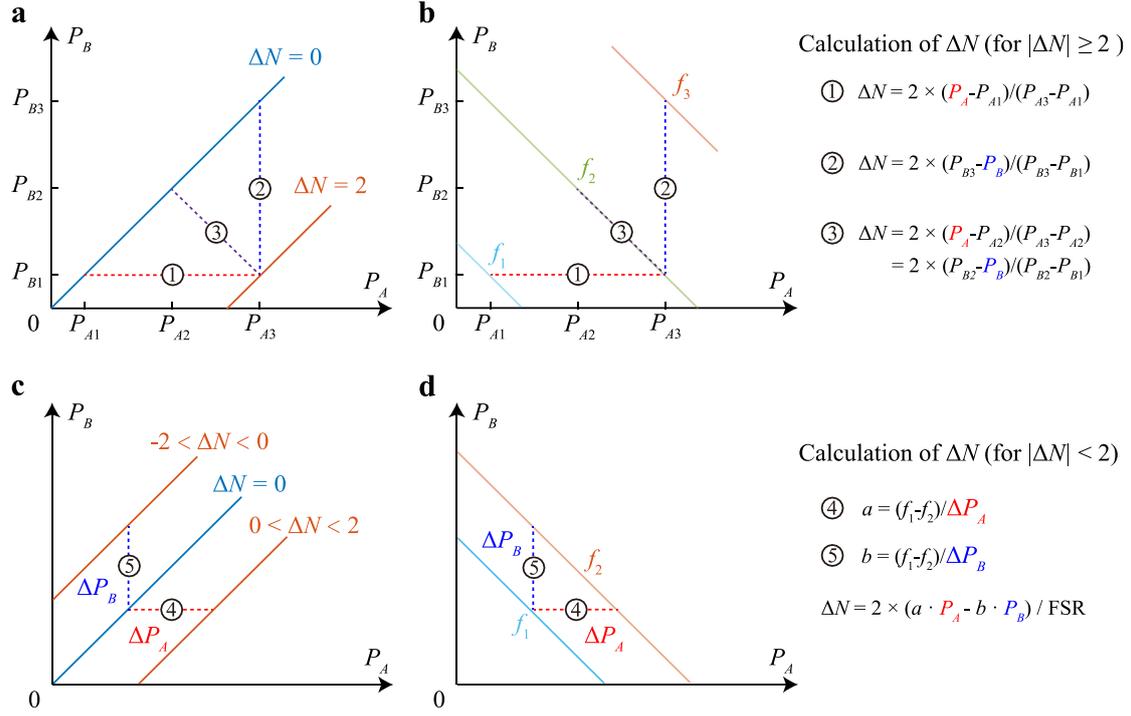

**Fig. S5** Characterization of the effective spatial modulation frequency detuning $\Delta N$ for **(a-b)** $|\Delta N|_{\max} \geq 2$ and **(c-d)** $|\Delta N|_{max} < 2$.

## Section 3. Numerical simulation of tunable µOPOs

In this section, we numerically investigate the nonlinear dynamics of the tunable OPO using the coupled mode equation (CME) model[2]:

$$\frac{d\tilde{A}_\mu}{dt} = -iD_{int}(\mu)\tilde{A}_\mu - \left(\frac{\kappa}{2} + i\alpha\right)\tilde{A}_\mu + ig \sum_{\mu_1+\mu_2-\mu_3=\mu} \tilde{A}_{\mu_1}\tilde{A}_{\mu_2}\tilde{A}^*_{\mu_3} + \delta_{\mu,0}\sqrt{\kappa_c}s_{in} \quad (S16)$$

where $\tilde{A}_\mu(t)$ denotes the time-varying amplitude of the resonant mode $\mu$, incorporating all frequency variations and deviation of the modes from the equidistance grid with spacing $D_1 = 2\pi \cdot \text{FSR}$. The cavity decay rate $\kappa = \kappa_i + \kappa_c$ represents the total losses, including the intrinsic losses $\kappa_i$ and the coupling rate from cavity to the bus waveguide $\kappa_c$. The term $\alpha = \omega_0 - \omega_p$ describes the detuning between the angular frequencies of the pumped resonance $\omega_0$ and the continue-wave pump laser frequency $\omega_p$. Here, the driving term $|s_{in}|^2 = P_{in}/\hbar\omega_0$ represents the photon flux of the pump, while the nonlinear term $g = \hbar\omega_0^2 c n_2/n_0^2 V_{eff}$ denotes the Kerr shift per photon, where $c$ is the speed of light, $n_2$ is the nonlinear refractive index, $n_0$ is the effective group refractive index and $V_{eff}$ is the effective optical mode volume.

The global dispersion is introduced through the integrate dispersion $D_{int}(\mu) = \omega_\mu - \omega_0 - \mu D_1$, where $\text{FSR} = D_1/2\pi$ is the free spectrum range. In the reconfigurable PhCR with target mode $\mu = 1$, the dispersion is modified by the time-varying mode coupling rate $\beta(\mu, t)$:

$$D_{int}(\mu, t) = \frac{1}{2}D_2\mu^2 + o(\mu^2) + \beta(\mu, t) - \beta(0, t) \quad (S17)$$

$$\beta(\mu, t) = \frac{\beta_0}{2}\text{sinc}(\mu - 1 + \Delta N/2) \quad (S18)$$

The simulation results are based on the split-step Fourier method and the CME model mentioned above. The simulations are performed with 512 modes, and the dispersion alterations $\beta_m$ are applied to the integrated dispersion $D_{int}(\mu)$. We used the typical parameters of our devices in the simulations: FSR = 300 GHz, $D_2/2\pi = 7.6$ MHz, $Q_i = 1.3 \times 10^6$, $Q_c = 0.65 \times 10^6$, $\omega_0/2\pi = 192.17$ THz, $\alpha/\kappa = 0.75$, $P_{in} = 200$ mW. Higher-order dispersion ($D_{n>2}$), Raman and thermal effects are omitted.

By applying time-varying phase shift $\beta(0, t)$ at the pump mode without considerating of adjacent mode splittings, we present the evolution of the OPO output spectra in **Fig. S6a**. As the detuning $\Delta N$ is swept from 2 to zero, the output signal and idler modes are dynamically tuned apart from each other, with conversion efficiency ranging from 0.2 to 0.4 (**Fig. S6b**). **Figure S6c** displays the stabilized OPO output spectra for several dispersion alterations as dictated in Fig. S6a and S6b, following 1,000 roundtrips for the system to reach a stable state.

Furthermore, we investigate the non-zero adjacent mode splittings during the tuning process, which slightly alter the local dispersion profiles around the pump mode. By repeating the simulation with a broadband, time-varied $\beta(\mu, t)$ defined by Eq. S18, we present the evolution of the OPO in **Fig. S6d** and **S6e**. The OPO output exhibits a similar tuning mechanism to that of the ideal dispersion tuning (Fig. S6a-c), with only a slight perturbation observed in the sweeping traces as the detuning ΔN is swept between 1.1 to 0.7. This slight variation is attributed to mode competition induced by the non-ideal adjacent mode splittings. However, we demonstrate that the available conversion efficiency does not deteriorate significantly when the OPO enters a stable state (**Fig. S6f**).

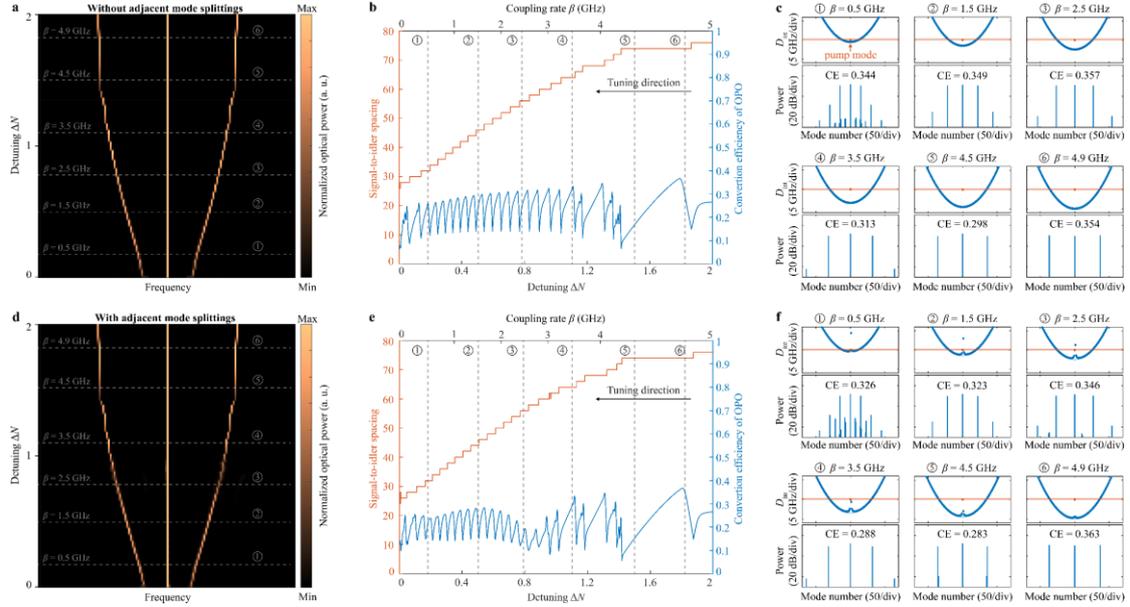

**Fig. S6** Numerical simulation of the tunable OPO with (**a-c**) and without (**d-f**) adjacent mode splittings.

## Section 4. Table of device parameters

Table 1. Device parameters.

| Data set | Ring height | Ring width (mean) | Ring radius (mean) | Grating pitch | Grating amplitude | Target mode $m_0$ | Mode splitting |
|---|---|---|---|---|---|---|---|
| Fig. 2b | ≈850 nm | ≈2.2 μm | 80 μm | ≈861 nm ($\theta = \pi$) | ≈60 nm ($\theta = \pi$) | 576 ($\theta = \pi$) | 5.56 GHz ($\theta = \pi$) |
| Fig. 2c-d | ≈750 nm | ≈2.3 μm | 80 μm | ≈862 nm | ≈45 nm | 575 | 3.85 GHz ($\theta = 2\pi$) |
| Fig. 3c-f | ≈650 nm | ≈2.25 μm | 80 μm | ≈877 nm | ≈90 nm | 565 | 7.60 GHz |
| Fig. 4-5 | ≈850 nm | ≈2.2 μm | 80 μm | ≈862 nm | ≈100 nm | 575 | 10.42 GHz |

## Supplementary References